\documentclass[submission, Proceedings]{SciPost}
\usepackage{amssymb,epsf,epsfig,amsmath,color}
\usepackage{multirow}
\usepackage{accents}
\usepackage{braket}
\usepackage{float}

\begin{document}

\begin{center}{\Large \textbf{
Probing Lepton Universality with (Semi)-Leptonic \bf{$B$} decays
}}\end{center}

\begin{center}
G. Banelli\textsuperscript{1},
R. Fleischer\textsuperscript{1,2},
R. Jaarsma\textsuperscript{1},
G. Tetlalmatzi-Xolocotzi\textsuperscript{1*}.
\end{center}

\begin{center}
{\bf 1} Nikhef, Science Park 105, NL-1098 XG Amsterdam, Netherlands\\
{\bf 2} Faculty of Science, Vrije Universiteit Amsterdam,\\
NL-1081 HV Amsterdam, Netherlands\\
* gtx@nikhef.nl
\end{center}

\begin{center}
\today
\end{center}

\definecolor{palegray}{gray}{0.95}
\begin{center}
\colorbox{palegray}{
  \begin{tabular}{rr}
  \begin{minipage}{0.05\textwidth}
    \includegraphics[width=8mm]{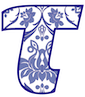}
  \end{minipage}
  &
  \begin{minipage}{0.82\textwidth}
    \begin{center}
    {\it Proceedings for the 15th International Workshop on Tau Lepton Physics,}\\
    {\it Amsterdam, The Netherlands, 24-28 September 2018} \\
    \href{https://scipost.org/SciPostPhysProc.1}{\small \sf scipost.org/SciPostPhysProc.Tau2018}\\
    \end{center}
  \end{minipage}
\end{tabular}
}
\end{center}


\section*{Abstract}
{\bf
The most recent measurements of the observables $R_{D^{(*)}}$ are in tension with the Standard Model offering hints of New Physics in $b\rightarrow c \ell \bar{\nu}_{\ell}$ transitions. Motivated by these results, in this work we present an analysis on their $b\rightarrow u \ell \bar{\nu}_{\ell}$ counterparts (for $\ell=e, ~\mu, ~\tau$). Our study has three main objectives. Firstly, using ratios of branching fractions, we assess the effects of beyond the Standard Model scalar and pseudoscalar particles in leptonic and semileptonic $B$ decays ($B^-\rightarrow \ell^- \bar{\nu}_{\ell}$,  $\bar{B}\rightarrow \pi  \ell \bar{\nu}_{\ell}$ and $\bar{B}\rightarrow \rho  \ell \bar{\nu}_{\ell}$). Here a key role is played by the leptonic $B$ processes, which are highly sensitive to new pseudoscalar interactions. In particular, we take advantage of the most recent measurement of the branching fraction of the channel $B^-\rightarrow \mu^-\bar{\nu}_{\mu}$  by the Belle collaboration. Secondly, we extract the CKM matrix element $|V_{ub}|$ while accounting simultaneously for New Physics contributions. Finally, we provide predictions for the branching fractions of yet unmeasured leptonic and semileptonic $B$ decays.
}

\vspace{10pt}
\noindent\rule{\textwidth}{1pt}
\tableofcontents\thispagestyle{fancy}
\noindent\rule{\textwidth}{1pt}
\section{Introduction}
\label{sec:intro}
Recent measurements of the observables $R_{D^{^{(*)}}}\equiv \mathcal{B}(B\rightarrow D^{(*)}\tau \bar{\nu}_{\tau})/\mathcal{B}(B\rightarrow D^{(*)}\ell' \bar{\nu}_{\tau})$, with $\ell'= e,~\mu$, have caused a lot of excitement in the high-energy physics community. As a matter of fact, the combined measurements of the BaBar, LHCb and Belle collaborations show a $3.9~\sigma$ deviation with respect to the expected value from the Standard Model (SM) \cite{Amhis:2016xyh}. If this effect is confirmed by forthcoming experimental determinations, it will indicate the presence of New Physics (NP) in the
exclusive decays $\bar{B}\rightarrow D^{(*)}\tau \bar{\nu}_{\tau}$, which are caused by the quark-level transition $b\rightarrow c \tau \bar{\nu}_{\tau}$ \cite{Svjetlana:2018}. Motivated by these results, we investigate the presence of NP effects in  $b\rightarrow u \ell \bar{\nu}_{\ell}$ processes, where $\ell=e,~\mu,~\tau$. To derive the relevant constraints, our study involves the interplay of the leptonic decay channels $B^-\rightarrow \ell^- \bar{\nu}_{\ell}$ with the semileptonic transitions $\bar{B}\rightarrow \pi \ell \bar{\nu}_{\ell}$ (analogous to $\bar{B}\rightarrow D \ell \bar{\nu}_{\ell}$) and  $\bar{B}\rightarrow \rho \ell \bar{\nu}_{\ell}$ (analogous to $\bar{B}\rightarrow D^{*} \ell \bar{\nu}_{\ell}$).\\

\noindent
The different decays to be included in this study are sensitive to NP scalar, pseudoscalar, vector and tensor interactions \cite{Hou:1992sy, Sakaki:2013bfa}. The pseudoscalar components are special because, due to the structure of the equations for the branching fractions $\mathcal{B}(B^-\rightarrow \ell^{-}\bar{\nu}_{\ell})$,
they lift the helicity suppression appearing in the corresponding SM expressions. This effect leads to interesting phenomenological predictions. Therefore,  we focus on the pseudoscalar contributions and complement our NP analysis by including also scalar operators, which are their natural partners in terms of the Lorenz structure.\\

\noindent
We follow an effective theory approach, with the low-energy Hamiltonian 

\begin{equation}\label{Heff}
{\cal H}_{\rm eff}= \frac{4 G_{\rm F}}{\sqrt{2}}V_{ub}\left[C_{V_L} {\cal O}_{V_L}^\ell  +
C_{S}^\ell{\cal O}_{S}^\ell   
+ 
C_{P}^\ell{\cal O}_{P}^\ell  \right] + \hbox{h.c.},
\end{equation}

\noindent
where

\begin{eqnarray}
\mathcal{O}^{\ell}_{V_L}=(\bar q \gamma^\mu P_L b)(\bar \ell \gamma_\mu P_L  \nu_{\ell}),
\quad
{\cal O}^{\ell}_{S}= (\bar q b)(\bar \ell   P_L \nu_{\ell}),\quad 
{\cal O}^{\ell}_{P}= (\bar q \gamma_5  b)(\bar \ell P_L  \nu_{\ell}),
\end{eqnarray}

\noindent
are the corresponding vector, scalar and pseudoscalar operators, respectively. 
The short-distance contributions are encoded in the Wilson coefficients $C_{V_L}$, $C^{\ell}_S$ and $C^{\ell}_P$. In the SM, only $C_{V_L}$ is non-vanishing and takes the value $C_{V_L}=1$. However, $C^{\ell}_S$ and $C^{\ell}_P$ may be non-zero in NP scenarios. An important example is the Two Higgs Doublet Model (2HDM) \cite{Hou:1992sy}, where the scalar and pseudoscalar coefficients are related to each other as follows:

\begin{eqnarray}
C^{\ell}_{S}&=& C_P^\ell = -\tan^2\beta \Bigl(\frac{m_b m_l}{M^2_{H^{\pm}}}\Bigl).
\end{eqnarray}

\noindent 
In general, the Wilson coefficients in Eq.~(\ref{Heff}) can be complex, with CP-violating NP phases. Unfortunately, direct CP asymmetries associated with leptonic and semileptonic processes cannot be used in their determination, because they vanish due to the absence of sizeable CP conserving phase differences. Hence, as discussed in Ref. \cite{Banelli:2018fnx}, in order to search for new sources of CP violation, a different strategy based on correlations between magnitudes and phases of Wilson coefficients should be followed. For simplicity, our discussion  in this report will be limited to real $C^{\ell}_S$ and $C^{\ell}_P$.
\\

\noindent
It is important to bear in mind that the exclusive determination of the CKM matrix element $|V_{ub}|$ is done using semileptonic $B$ decays originating from the quark level transition $b\rightarrow u \ell \bar{\nu}_{\ell}$ assuming the SM. However, this value may be affected by NP effects. Therefore,  in order to take into account these contributions we propose the following strategy:

\begin{enumerate}
	\item Using combinations of ratios of branching fractions for 
	 leptonic and semileptonic $B$ decays, where $|V_{ub}|$ cancels, we determine the allowed regions for $C^{\ell}_S$ and $C^{\ell}_P$.
	
	\item We then use these numerical ranges for 
	$C^{\ell}_S$ and $C^{\ell}_P$ and evaluate the branching fraction  for given leptonic or semileptonic $B$ processes, allowing us to finally extract  $|V_{ub}|$ from data.
\end{enumerate}

\noindent
In the following sections, we will elaborate on this procedure and we will present our predictions for the branching  fractions of leptonic and semileptonic $B$ decays which have not yet been measured. The full study can be found in Ref.~\cite{Banelli:2018fnx}.

\section{Constraints from leptonic $	B$ decays}

In the SM, the branching fraction of the process $B^{-}\rightarrow \ell^{-}\bar{\nu}_{\ell}$ is given by

\begin{equation}\label{SM-Br}
{\mathcal B}(B^-\to\ell^-\bar\nu_\ell)|_{\rm SM}=\frac{G_{\rm F}^2}{8\pi}
|V_{ub}|^2M_{B^-}m_\ell^2\left(1-\frac{m_\ell^2}{M_{B^-}^2}\right)^2f_{B^-}^2\tau_{B^-}.
\end{equation}
 
\noindent
 We observe that the branching ratio is proportional to the square of the mass of the lepton in the final state $m^2_{\ell}$. For electrons and muons, this factor very strongly suppresses the decay probabilities, which is referred as helicity suppression. Interestingly, it is also present in the rare  decays $B_{d,s}\rightarrow \ell^+\ell^-$~ \cite{Fleischer:2017ltw}. Since $m_{e}, m_{\mu} \ll m_{\tau}$, the helicity suppression is less effective for tau leptons.\\

\noindent
Leptonic $B$ decays are very clean channels, where all the non-perturbative hadronic information is encoded in the decay constant given by \cite{Aoki:2016frl, Dowdall:2013tga}
\begin{eqnarray}
f_{B^-}=0.186\pm 0.004.
\end{eqnarray}

\noindent
Assuming the SM, we use the CKMFitter value \cite{Charles:2004jd}
\begin{eqnarray}
|V_{ub}|=(3.601\pm 0.098)\times 10^{-3},
\label{eq:fBs}
\end{eqnarray}

\noindent
together with Eq.~(\ref{eq:fBs}) to obtain
\begin{eqnarray}
{\mathcal B}(B^-\to\tau^-\bar\nu_\tau) &=& (7.92\pm0.55)\times10^{-5}, \nonumber\\
{\mathcal B}(B^-\to\mu^-\bar\nu_\mu) &=& (3.56\pm0.25)\times10^{-7}, \label{eq:SMleptBrmu} \nonumber\\
{\mathcal B}(B^-\to e^-\bar\nu_e) &=& (8.33\pm0.58)\times10^{-12},\label{eq:SMleptBr}
\end{eqnarray}

\noindent
where due the tiny value of the mass of the electron, the helicity suppression leads to a extremely small branching 
fraction for the channel $B^-\to e^-\bar\nu_e$.\\

\noindent
The experimental results reported by BaBar, Belle and LHCb lead to
\begin{eqnarray}\label{eq:leptonicExp}
{\mathcal B}(B^-\to\tau^-\bar\nu_\tau)&=& (1.09\pm0.24)\times10^{-4},\nonumber\hbox{\cite{PhysRevD.98.030001}}\\
{\mathcal B}(B^-\to\mu^- \bar\nu_\mu)&=& (6.46 \pm 2.74)\times10^{-7},\hbox{\cite{Sibidanov:2017vph}}\nonumber\\
{\mathcal B}(B^- \to e^- \bar\nu_e) &<& 9.8 \times 10^{-7} \, \mbox{(90\% C.L.)}\hbox{\cite{Satoyama:2006xn}}.
\end{eqnarray}

\noindent
Here the measurement corresponding to the channel $B^-\to\mu^- \bar\nu_\mu$ was reported recently by the Belle  collaboration with a $2.4~\sigma$ excess over background and will be key during our phenomenological study.\\
 
 \noindent
 For leptonic $B$ decays, the hadronic matrix element of the $O_S$ operator  in Eq.~(\ref{Heff}) vanishes. Consequently, is does not receive scalar NP contributions.
 Once pseudoscalar NP effects are taken into account, Eq.~(\ref{SM-Br}) gets modified as
 
 \begin{eqnarray}
 {\mathcal B}(B^-\to\ell^-\bar\nu_\ell)={\mathcal B}(B^-\to\ell^-\bar\nu_\ell)|_{\rm SM}
 \left|1+ \frac{M_{B^-}^2}{m_\ell (m_b+m_u)} C_P^\ell \right|^2.
 \end{eqnarray}

\noindent  
In the case of electrons and muons, the helicity suppression is lifted by the  mass ratio

\begin{eqnarray}
M^2_{B^-}/\Bigl[m_{\ell} (m_b +m_u)\Bigl]\sim M_{B^-}/m_{\ell},
\end{eqnarray}

\noindent
thereby amplifying the effects of $C_P^\ell$.\\
  
\begin{center}
	\begin{figure}[H]
		\begin{center}
			\includegraphics[width=0.6\textwidth]{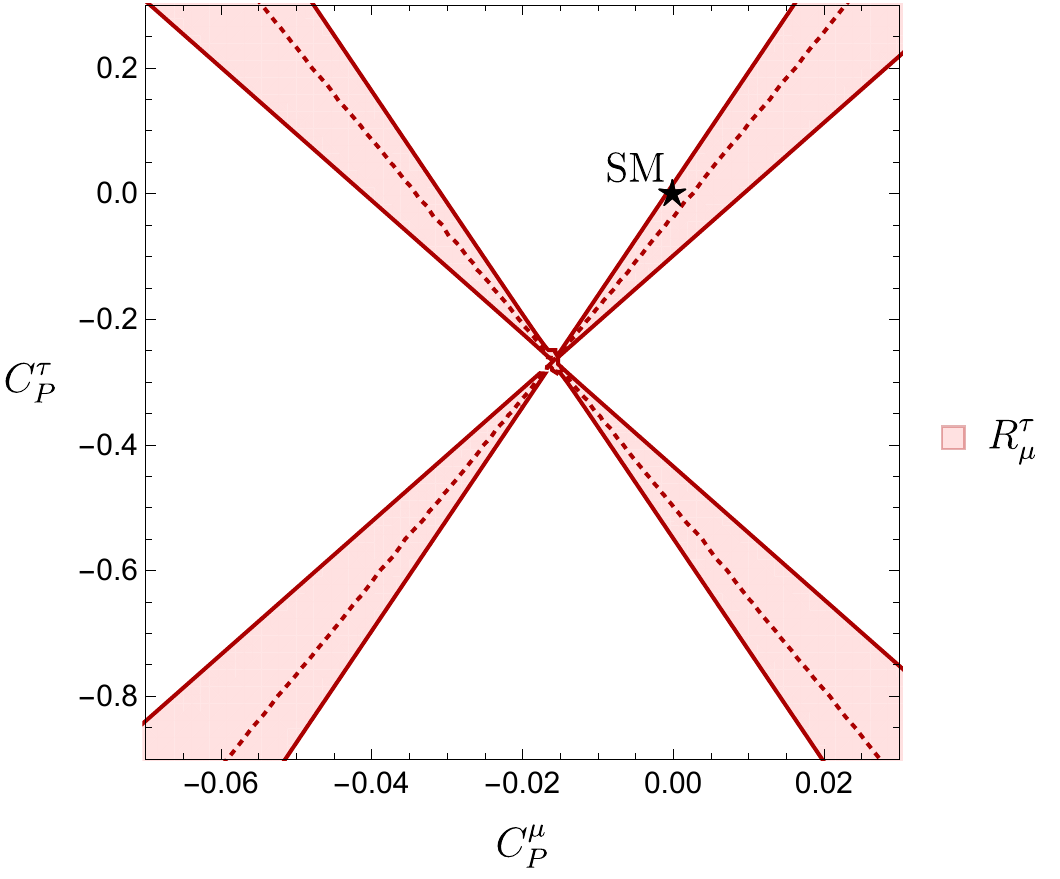} \qquad
			\caption{Allowed regions in the  $C^{\mu}_P$--$C^{\tau}_P$ plane following from the leptonic ratio $R^{\tau}_{\mu}$.}
			\label{fig:a}
		\end{center}
	\end{figure}
\end{center}

\noindent  
To constrain the pseudoscalar Wilson coefficients $C^{\mu}_P$ and $C^{\tau}_P$, we consider the observable 

\begin{eqnarray}\label{eq:leptonicratio}
R^{\tau}_{\mu}\propto 
\frac{{\mathcal B}(B^-\to\tau^-\bar\nu_{\tau})}{{\mathcal B}(B^-\to \mu^-\bar\nu_{\mu})},
\end{eqnarray}

\noindent
where the normalization factor is chosen in such a way that we get $R^{\tau}_{\mu}=1$ in the SM. The main features of $R^{\tau}_{\mu}$  are the cancellation of the hadronic decay constant $f_{B^-}$ and of the CKM matrix element $|V_{ub}|$. By comparing the corresponding theoretical determination for this ratio with the experimental result, we obtain the regions shown in Fig.~\ref{fig:a}. Here we can see that, even though $R^{\tau}_{\mu}$ is already imposing strong constraints on the values that $C_P^\mu$ and $C_P^\tau$ can take, the arms of the resulting cross-shaped area extend to infinity.  To improve our bounds on these pseudoscalar Wilson coefficients, we have to include more observables sensitive to $C^{\mu}_P$ and $C^{\tau}_P$. This topic will be discussed in the next section.

\section{Semileptonic \bf{$B$} decays}

To derive stronger constraints on $C^{\mu}_P$ and $C^{\tau}_P$,  we consider the branching ratio $\mathcal{B}(\bar{B}\rightarrow \rho \ell \bar{\nu}_{\ell})$, which in analogy with the leptonic decays, does not depend on scalar NP contributions \cite{Sibidanov:2013rkk}. Due to the presence of the $\rho$ meson in the final state, the hadronic contributions are more complicated than those encountered in the leptonic $B$ decays. In the helicity basis, these effects are encoded in the form factors denoted as $H^{\rho}_{V,+}$, $H^{\rho}_{V,-}$, $H^{\rho}_{V,0}$, $H^{\rho}_{V,t}$ and $H^{\rho}_{S}$. The non-perturbative technique employed for their calculation depends on the value of the square of the four-momentum transferred $q^2$ to $\ell$ and $\bar{\nu}_{\ell}$.  In the literature, two main approaches are usually considered for their determination:

\begin{itemize}
	\item QCD sum rules for the low energy regime $0\leq q^2\leq q^2_{\rm max}$, where typically $q^2_{\rm max}\in  [12, 16]~\rm{GeV}^2$.
	
	\item Lattice QCD calculations \cite{Lattice:2015tia, Bowler:2004zb} are applied when $q^2$ is close to the maximal leptonic momentum transfer: $q^2_{\rm max}\leq q^2\leq (M_{B}-M_{\rho})^2$. 	
\end{itemize}

\noindent
As experimental input we consider the measurements of
${\mathcal B}(\bar B^0\rightarrow \rho^+ \ell^- \bar{\nu}_{\ell})$ and ${\mathcal B}(B^-\rightarrow \rho^0 \ell^- \bar{\nu}_{\ell})$  reported by Belle in 2013, which include an admixture of electrons and  muons in the final state \cite{Sibidanov:2013rkk}.  Using the isospin symmetry, we combine these two measurements to obtain  $\Braket{{\mathcal B}(\bar B\rightarrow \rho \ell^- \bar{\nu}_{\ell})}_{[\ell=~ e, \mu],~q^2\leq 12~\rm{GeV}^2}=(1.98 \pm 0.12) \times 10^{-4}$.  We introduce the ratio

\begin{eqnarray}\label{eq:leptonicoversemileptonicrho-theo}
\mathcal{R}^{\mu}_{\Braket{e, \mu}; \rho ~ [q^2\leq 12]~\rm{GeV}^2}&\equiv&\mathcal{B}(B^-\rightarrow \mu^- \bar{\nu})/
\Braket{{\mathcal B}(\bar B \rightarrow \rho \ell^- \bar{\nu}_{\ell})}_{[\ell=~ e, \mu],~q^2\leq 12~\rm{GeV}^2},
\end{eqnarray}
\noindent
where the CKM element $|V_{ub}|$ cancels.\\

\begin{center}
	\begin{figure}[H]
		\begin{center}
			\includegraphics[width=0.6\textwidth]{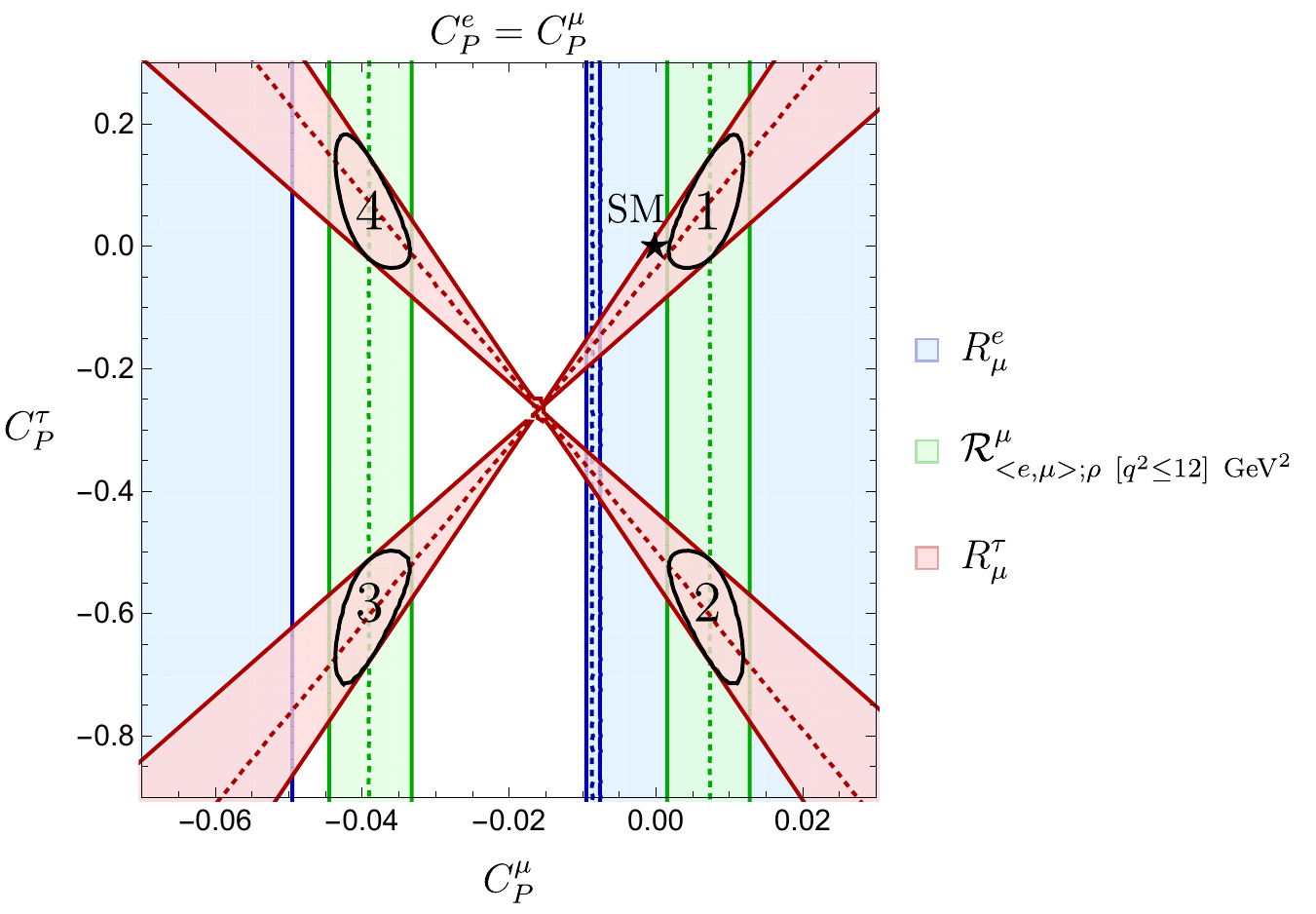} \qquad
			\caption{Allowed regions in the  $C^{\mu}_P$--$C^{\tau}_P$ plane utilizing the ratios $R^{e}_{\mu}$, $R^{\tau}_{\mu}$ and $\mathcal{R}^{\mu}_{\Braket{e, \mu}; \rho~[q^2\leq 12]~\rm{GeV}^2}$ under the assumption $C^e_P=C^{\mu}_P$.}
			\label{fig:lepsemmutau}
		\end{center}
	\end{figure}
\end{center}

\noindent
Since the experimental determinations do not yet provide independent information for electrons and muons, we can only  obtain the allowed values for $C^{e}_P$ and $C^{\mu}_P$ if we correlate these Wilson coefficients through different assumptions. We start by testing the  hypothesis of having
 universal NP interactions in electrons and muons, i.e.~$C^{e}_P=C^{\mu}_P$, and explore the  behaviour of the semileptonic decay in the range $q^2\leq 12~\rm{GeV}^2$, where an analytical parameterization from QCD sum rules is available \cite{Straub:2015ica}. Then, we proceed to determine the allowed regions in the $C^{\mu}_P$ $-$ $C^{\tau}_P$  plane  by using the observables $R^{\tau}_{\mu}$ and
 $\mathcal{R}^{\mu}_{\Braket{e, \mu}; \rho ~ [q^2\leq 12]~\rm{GeV}^2}$ as constraints . Moreover, we include the leptonic ratio $R^{e}_{\mu}$, which is analogous to $R^{\tau}_{\mu}$ in Eq.~(\ref{eq:leptonicratio}),  calculated from the experimental bound available for $\mathcal{B}(B^-\rightarrow e^- \bar{\nu}_e)$ presented in Eq.~(\ref{eq:leptonicExp}). The resulting plot is shown in Fig.~\ref{fig:lepsemmutau}, where only the elliptical areas labelled as ``1'' and ``2'' are allowed. Even though solution  ``1'' is  compatible with the SM, the solution inside  region  ``2'', corresponding to NP, is not excluded.

\noindent
So far, our treatment has been focused on constraining  the pseudoscalar Wilson coefficient $C^{\ell}_P$. In order to have sensitivity  on the scalar Wilson coefficient $C^{\ell}_S$, we include the branching ratios of the processes $\bar B\rightarrow \pi \ell \bar{\nu}_{\ell}$. 
Unfortunately, the current experimental information does not provide independent measurements for electrons and muons. To incorporate these processes in our analysis, we make an average using the isospin symmetry to combine 
the independent contributions of $B^0$ and $B^-$ provided in Ref. \cite{PhysRevD.98.030001}, yielding $\Braket{{\mathcal B}(\bar{B}\rightarrow \pi \ell \bar{\nu}_{\ell})}_{[\ell=~ e, \mu]}=(1.53 \pm 0.04)\times 10^{-4}$. \\

\noindent
Two more observables, which are sensitive to both $C^{\ell}_S$  and $C^{\ell}_P$, are now at our disposal:

\begin{eqnarray}\label{eq:leptonicoversemileptonicpi}
\mathcal{R}^{\mu}_{\Braket{e, \mu}; \pi}&\equiv&\mathcal{B}(B^-\rightarrow \mu^- \bar{\nu})/
\Braket{{\mathcal B}(\bar B \rightarrow \pi \ell^- \bar{\nu}_{\ell})},\nonumber\\
\mathcal{R}^{\Braket{e, \mu};\rho~[q^2_{\rm{min}} \leq q^2 \leq q^2_{\rm{max}}]}_{\Braket{e, \mu};\pi}
&\equiv&
\Braket{ \mathcal{B}(\bar{B}\rightarrow \rho \ell^- \bar{\nu}_{\ell})}_{[\ell=e, \mu]}\Bigl|^{q^2_{\rm{max}}}_{q^2_{\rm{min}}}/
\Braket{\mathcal{B}(\bar{B}\rightarrow \pi \ell^- \bar{\nu}_{\ell})}_{[\ell=e, \mu]}.
\label{eq:semirhosemipi}
\end{eqnarray}

\begin{center}
	\begin{figure}
		\begin{center}
			\includegraphics[width=0.65\textwidth]{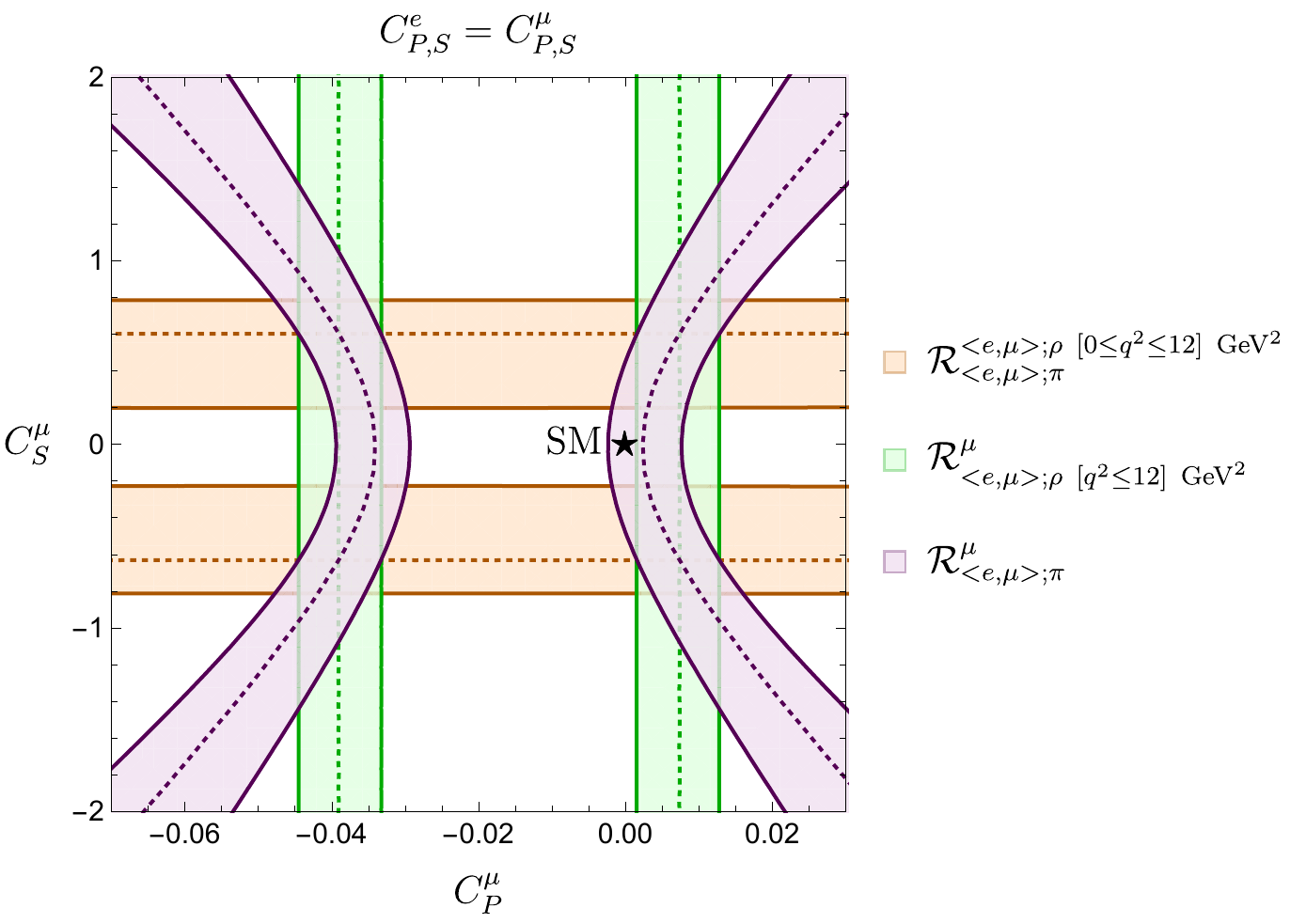} \qquad
			\caption{The allowed regions in the  $C^{\mu}_P$--$C^{\mu}_S$ plane following from the observables
				$\mathcal{R}^{\Braket{e, \mu};\rho~[0 \leq q^2 \leq 12]~{\rm GeV^2}}_{\Braket{e, \mu};\pi}$, $\mathcal{R}^{\mu}_{\Braket{e, \mu};\rho~[q^2 \leq 12]~{\rm GeV^2}}$ and $\mathcal{R}^{\mu}_{\Braket{e, \mu};\pi}$.}
			\label{fig:semilepoversemilep}
		\end{center}
	\end{figure}
\end{center}

\noindent
Including furthermore $\mathcal{R}^{\mu}_{\Braket{e, \mu}; \rho ~ [q^2\leq 12]~\rm{GeV}^2}$, introduced in Eq.~(\ref{eq:leptonicoversemileptonicrho-theo}), and making the assumptions  $C^e_P=C^{\mu}_P$, $C^e_S=C^{\mu}_S$, we obtain the regions in the $C^{\mu}_S$$-$$C^{\mu}_P$  plane  shown in Fig.~\ref{fig:semilepoversemilep}. We observe that $\mathcal{R}^{\Braket{e, \mu};\rho~[0 \leq q^2 \leq 12]~\rm{GeV}^2}_{\Braket{e, \mu};\pi}$ results in two horizontal bands, which are in tension with the SM at $(1$$-$$2)~\sigma$. 

\noindent
We have restricted our studies to the low $q^2$ regime. Let us now investigate whether the tension found with the SM persists for large $q^2$ values. 
To the best of our knowledge, the only available hadronic inputs for $B\rightarrow \rho$ transitions in the $12~\hbox{GeV}^2 \leq q^2$ range come from lattice determinations \cite{Bowler:2004zb} and were obtained in 2004. Due to the unstable nature of the $\rho$ meson, these  calculations are rather challenging. To keep the non-perturbative uncertainties under control when evaluating the branching ratios, we use differential distributions in $q^2$ rather than fully  integrated expressions. Therefore, for $12~\hbox{GeV}^2<q^2$ we introduce the following observable:

\begin{eqnarray}\label{eq:dR}
d\mathcal{R}^{\Braket{e, \mu};\rho}_{\Braket{e, \mu};\pi}&=&
\frac{2\Braket{d\mathcal{B}(B^-\rightarrow \rho^0 \ell^{-}\bar{\nu}_\ell)/dq^2}_{[\ell=e, \mu]}}{\Braket{\mathcal{B}(\bar{B}\rightarrow \pi \ell^- \bar{\nu}_{\ell})}_{[\ell=e, \mu]}}.
\end{eqnarray}

\begin{center}
	\begin{figure}
		\begin{center}
			\includegraphics[width=0.45\textwidth]{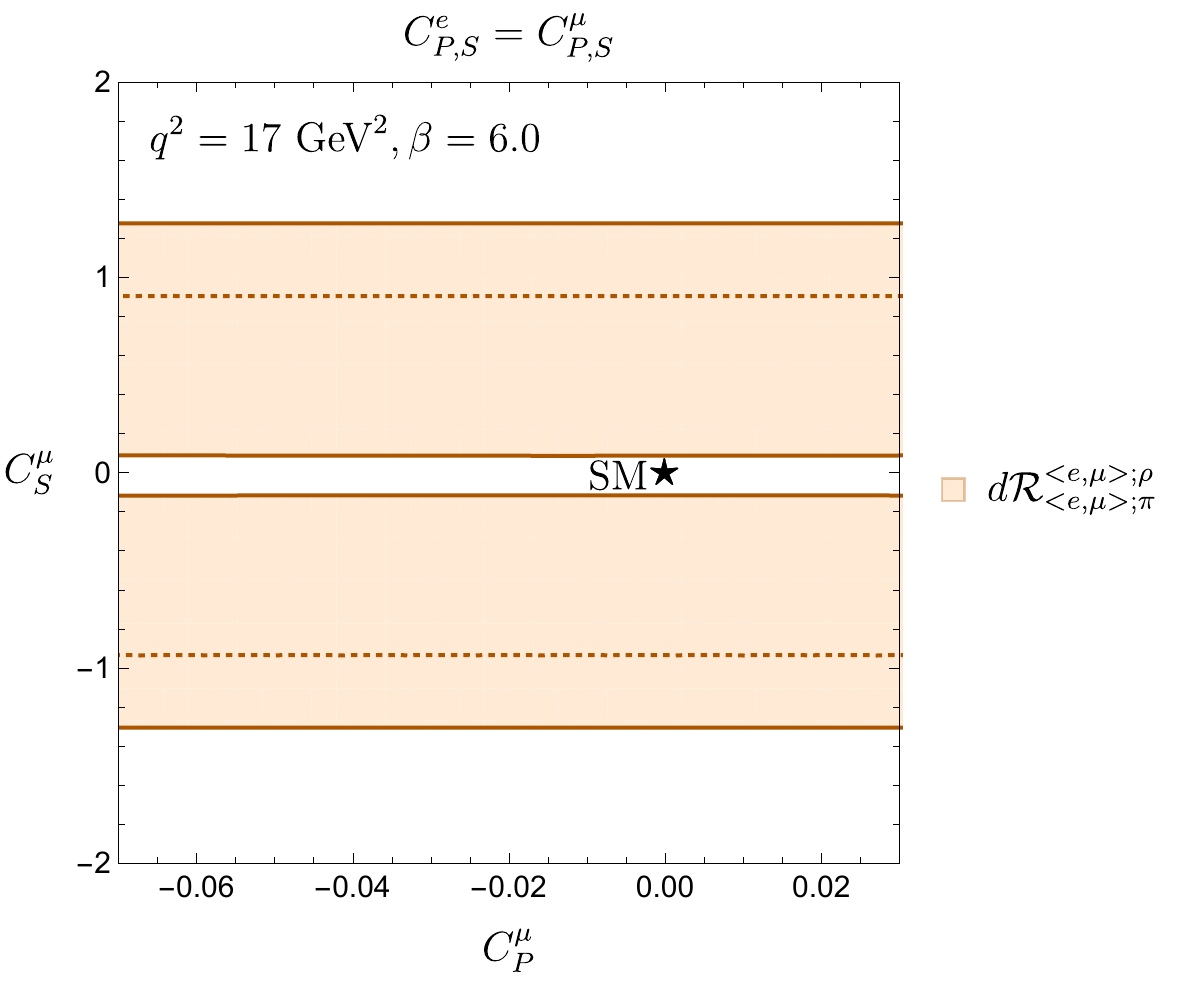}
			\caption{Regions in the $C^{\mu}_P$--$C^{\mu}_S$ plane following from the observable $d\mathcal{R}^{\Braket{e, \mu};\rho}_{\Braket{e, \mu};\pi}$ 
				in the large $q^2$ regime for  $q^2=17~\rm{GeV}^2$.}
			\label{fig:dsemilepoversemilepbinbybinbeta6}
		\end{center}
	\end{figure}
\end{center}

\noindent
Interestingly, as can be seen in Fig.~\ref{fig:dsemilepoversemilepbinbybinbeta6},  we find a mild tension with the SM for $q^2=17~\rm{GeV}^2$. To shed light on the origin of this feature, an updated determination of the form factors for the transition $B\rightarrow \rho$ in the high $q^2$ regime is needed. 

\section{Determination of \bf{$|V_{ub}|$} and predictions of the branching fractions for \bf{$B^- \to e^- \overline{\nu}_e$} and \bf{$\overline{B}\to \rho \tau^- \overline{\nu}_\tau$}}
\label{Sec:Vubandpred}

Having the regions for $C^{\ell}_{P}$ and $C^{\ell}_{S}$ available, we can proceed with the extraction of $|V_{ub}|$. The strategy introduced at the end of Sec.~\ref{sec:intro} describes the basic ingredients required for the determination of this CKM matrix element.  A more refined procedure, which accounts for possible correlations between our observables, is discussed in  \cite{Banelli:2018fnx}. By applying this method, we obtain

\begin{equation} \label{eq:vubSolUniversal}
|V_{ub}| = (3.31 \pm 0.32) \times 10^{-3}.
\end{equation}

\noindent
Although in agreement with the  value reported by the CKMFitter collaboration \cite{Charles:2004jd}, presented in Eq.~(\ref{eq:fBs}), the uncertainty in Eq.~(\ref{eq:vubSolUniversal}) is three times bigger.  However,  our target here is only to illustrate the application of our method which does not assume the SM during the extraction of $|V_{ub}|$.  Future improvements in the precision of our observables will allow us to reduce the uncertainty in the CKM matrix element $|V_{ub}|$. \\

\noindent
Besides universal pseudoscalar NP interactions between electrons and muons, other possible scenarios can be considered.  We have  also explored the cases:  $C^e_P=(1/10)C^{\mu}_P$ and $C^e_P=10 C^{\mu}_P$. In addition, we included the 2HDM, where the pseudocalar Wilson coefficients for electrons and tau leptons are correlated with $C^{\mu}_P$ as

\begin{eqnarray}
C^e_P=\frac{m_e}{m_{\mu}}\times C^{\mu}_P,&&
C^{\tau}_P=\frac{m_{\tau}}{m_{\mu}} \times C^{\mu}_P.
\end{eqnarray}

\noindent
In these scenarios, our method leads to the value for $|V_{ub}|$  in Eq.~(\ref{eq:vubSolUniversal}).  Finally, we have considered the situation where NP affects only the 3rd generation of leptons, i.e. $C^{\tau}_P \neq 0$, while $C^e_P=C^{\mu}_P=0$. Here we obtain $|V_{ub}|=(4.85 \pm 1.03) \times 10^{-3}$, which is closer to the value of $|V_{ub}|$ following from inclusive determinations \cite{Charles:2004jd, Aglietti:2004fz, Aglietti:2006yb, Aglietti:2007ik, Lange:2005yw, Bosch:2004th, Bosch:2004cb, Andersen:2005mj, Gambino:2007rp}.\\

\begin{center}
	\begin{figure}[H]
		\begin{center}
			\includegraphics[width=0.6\textwidth]{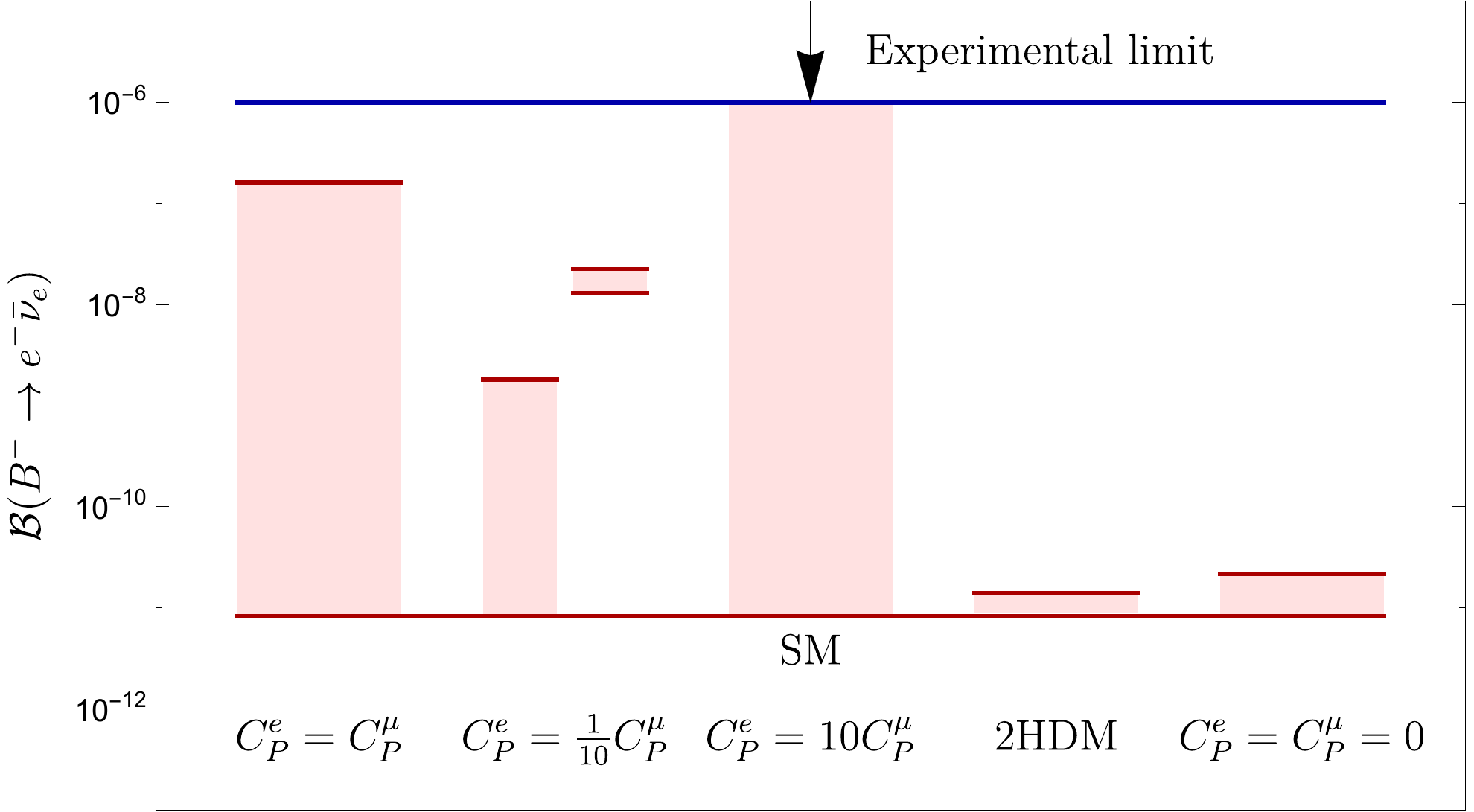}
			\caption{Illustration of the possible enhancement of  $\mathcal{B}(B^-\rightarrow e^- \bar{\nu}_{e})$ for the scenarios discussed in the text. The blue line gives the current experimental upper bound on $\mathcal{B}(B^-\rightarrow e^- \bar{\nu}_{e})$, whereas the red horizontal line on the bottom represents the SM value. The red regions indicate the values of the branching ratio that may be obtained.}
			\label{fig:prop-plot-e}
		\end{center}
	\end{figure}
\end{center}

\noindent
Finally, we use the ranges for the different Wilson coefficients to make predictions for the yet unmeasured branching ratios $\mathcal{B}(B^-\rightarrow e^- \bar{\nu}_{e})~~\hbox{and}~~
\mathcal{B}(\bar B\rightarrow \rho \tau \bar{\nu}_{\tau})$. In Fig.~\ref{fig:prop-plot-e}, the predictions for $\mathcal{B}(B^-\rightarrow e^- \bar{\nu}_e)$ are illustrated. We would like to highlight that, for $C^e_P=10 C^{\mu}_P$, our analysis leads to a potential enhancement of
$\mathcal{B}(B^-\rightarrow e^- \bar{\nu}_e)$ that may even saturate the current experimental bound. An interesting phenomenological prediction in the case $C^{e}_P=C^{\mu}_P$ is an enhancement by up to four orders of magnitude with respect to the SM value, thereby lying just a factor of 10  below the current experimental upper bound. An analogous effect in the case of $B_s \rightarrow e^+ e^-$ has been discussed in Ref. \cite{Fleischer:2017ltw}.\\

\begin{center}
	\begin{figure}[H]
		\begin{center}
			\includegraphics[width=1.0\textwidth]{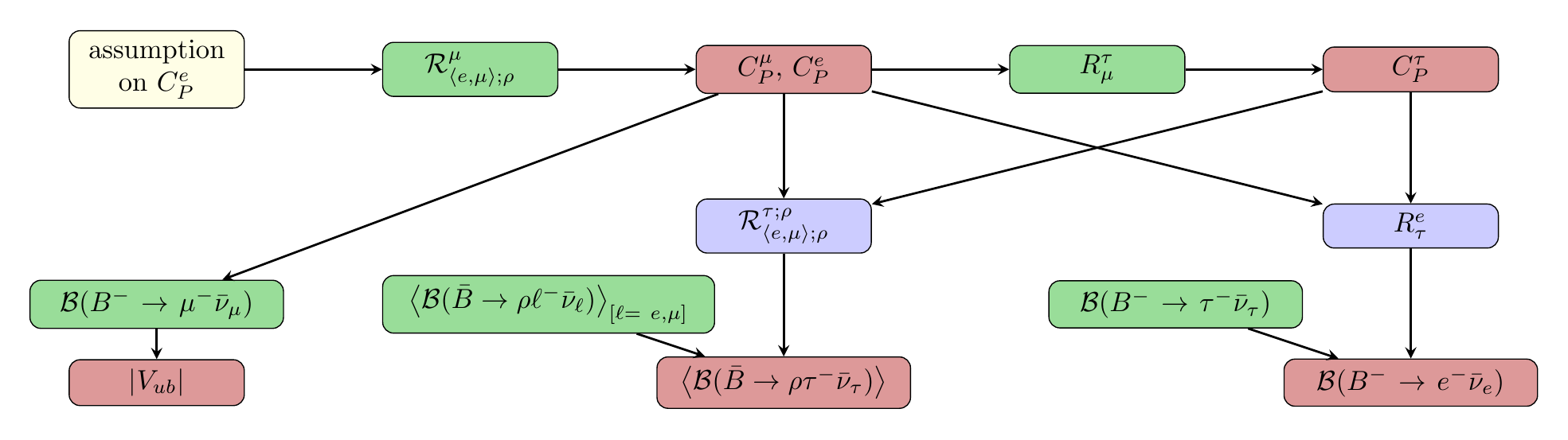}
			\caption{Flowchart illustrating our strategy.}
			\label{fig:flowchart}
		\end{center}
	\end{figure}
\end{center}

\noindent
The values for $\mathcal{B}(\bar B\rightarrow \rho \tau^- \bar{\nu}_{\tau})$ following from our analysis are  consistent with the SM picture at the $1~\sigma$ level. As discussed in \cite{Banelli:2018fnx}, future measurements of this observable are a powerful tool to distinguish between solutions ``1''  and ``2'' in Fig.~\ref{fig:lepsemmutau}. Our full strategy is summarized in the flow chart in Fig.~\ref{fig:flowchart}.

\vspace{1cm}

\newpage

\section{Outlook}
Leptonic and semileptonic  decays originating from $b\rightarrow u \ell \bar{\nu}_{\ell}$ transitions are very interesting and powerful channels to unveil potential NP contributions. They are the counterparts of the $b\rightarrow c \ell \bar{\nu}_{\ell}$ processes, where recent experimental results for the associated observables $R_{D^{(*)}}$ show effects which may originate from physics beyond the SM.\\

\noindent
We have presented a study which analyses the effects of scalar and pseudoscalar particles in leptonic and semileptonic decays arising from the transition
$b\rightarrow u \ell \bar{\nu}_{\ell}$. Central to our analysis is the high sensitivity of the leptonic $B$ decays to the presence of NP pseudoscalar contributions. To take advantage of this feature, we have used the most recent Belle measurement of the branching fraction of the channel $B^-\rightarrow \mu^- \bar{\nu}_{\mu}$. \\

\noindent
We have developed a strategy with three main goals. Firstly, we have obtained the corresponding short-distance  NP contributions,  utilizing ratios of leptonic and semileptonic processes where the CKM element $|V_{ub}|$ cancels. Secondly, we have determined the value of  $|V_{ub}|$, considering simultaneously the presence of NP effects. Finally, we have made predictions for branching fractions of non-yet measured $B$ decays.  In particular, we have addressed the decay channels $B^- \to e^- \bar{\nu}_e$ and $\bar{B}\to \rho \tau \bar{\nu}_\tau$, and found NP effects that may be within the reach of the Belle II collaboration and future updates of the LHCb experiment.\\

\noindent
 Our analysis includes the semileptonic processes $\bar{B}\rightarrow \pi \ell \bar{\nu}_{\ell}$ and $\bar{B}\rightarrow \rho \ell \bar{\nu}_{\ell}$ with $\ell=e,~\mu,~\tau$, where the current experimental data does not provide information on electrons and muons separately. In order  to test universality in light leptons of different flavours,  it would be desirable that experimental collaborations provide independent measurements for $\ell=e$  and $\ell=\mu$. In addition, a better understanding of the behaviour of the branching fraction of the process $\bar{B}\rightarrow \rho \ell \bar{\nu}_{\ell}$ requires an update of the corresponding non-perturbative contributions in the high $q^2$ regime. Following these lines we will be able to take further advantage of these semileptonic $B$ decays to search for NP. This will complement the ongoing searches in their counterparts
  $\bar{B}\rightarrow D^{(*)} \ell \bar{\nu}_{\ell}$, which enter the observables $R_{D^{(*)}}$.\\


\paragraph{Funding information}
This research project was supported
by the Netherlands Foundation for Fundamental Research of Matter (FOM) programme 156,
“Higgs as Probe and Portal”, and by the National Organisation for Scientific Research (NWO).




\bibliographystyle{SciPost_bibstyle}
\bibliography{SciPost_Example_BiBTeX_File}

\nolinenumbers

\end{document}